\documentclass[sigconf]{aamas}

\AtBeginDocument{%
  \providecommand\BibTeX{{%
    \normalfont B\kern-0.5em{\scshape i\kern-0.25em b}\kern-0.8em\TeX}}}

\usepackage{booktabs}    
\usepackage{flushend} 

\usepackage{amsmath, amsthm, latexsym, graphics, graphicx,amsfonts, color}
\usepackage{verbatim}

\usepackage{graphics,graphicx,adjustbox}
\usepackage{epsf}
\usepackage{stmaryrd}
\usepackage{longtable}
\usepackage{latexsym}
\usepackage{color}
\usepackage[mathscr]{eucal}
\usepackage[refpage,english]{nomencl}
\usepackage{xspace}
\usepackage{algorithmic} 
\usepackage{algorithm}
\usepackage{subfigure}
\usepackage{wrapfig}
\usepackage{url}
\usepackage{mathtools}
\usepackage{tikz}
\usepackage{balance}

\setcopyright{ifaamas}  
\copyrightyear{2020} 
\acmYear{2020}
\acmDOI{}
\acmPrice{} 
\acmISBN{} 
\acmConference[AAMAS'20]{Proc.\@ of the 19th International Conference on Autonomous Agents and Multiagent Systems (AAMAS 2020)}{May 9--13, 2020}{Auckland, New Zealand}{B.~An, N.~Yorke-Smith, A.~El~Fallah~Seghrouchni, G.~Sukthankar (eds.)}

\makeatletter
\DeclareRobustCommand*\cal{\@fontswitch\relax\mathcal}
\makeatother

\newcommand{\GA}{{\Gamma}} 

\let\none = -

\def\F {{\rm{F}}}

\def\F {{\rm F}}

\def\CTL {{${\rm CTL}$}}

\def\ATL {{${\rm ATL}$}}

\newcommand{\coop}[1]{\langle\!\langle{#1}\rangle\!\rangle}

\newcommand{\ATLSAT}[1]{${{\rm ATL}_{#1}{\rm SAT}}$}
\newcommand{\ATLSATi}{\ATLSAT{i}}
\newcommand{\ATLSATI}{\ATLSAT{I}}

\newcommand{\prop}[1]{\ensuremath{\mathsf{{#1}}}}
\newcommand{\Next}[1][]{X_{{#1}}  
        \,}
\newcommand{\Sometm}[1][]{F_{{#1}}}
\newcommand{\Always}[1][]{G_{{#1}}}
\newcommand{\Until}[1][]{U_{{#1}}}

\newcommand{\Epath}{\mathsf{E}}
\newcommand{\Apath}{\mathsf{A}}

\newcommand{\satisf}[1][]{\models}

\def\MsATL{MsATL\xspace} 

\def\paragraph2#1{\textit{#1.}}

\begin{document}

\title{\MsATL: a Tool for SAT-Based \ATL\ Satisfiability Checking}  
                     
\subtitle{Demonstration}

\author{Artur Niewiadomski}
\affiliation{
  \institution{Siedlce University, Faculty of Exact and Natural Sciences}
  }
\email{artur.niewiadomski@uph.edu.pl}

\author{Magdalena Kacprzak}
\affiliation{
  \institution{Bialystok University of Technology, Poland}
  }
 \email{m.kacprzak@pb.edu.pl}

\author{Damian Kurpiewski, Micha{\l} Knapik,\\ Wojciech Penczek}
\affiliation{
  \institution{Institute of Computer Science, Polish Academy of Sciences, Warsaw, Poland}
  }
\email{ {d.kurpiewski,m.knapik,penczek}@ipipan.waw.pl}

\author{Wojciech Jamroga}
\affiliation{
  \institution{Institute of Computer Science, Polish Academy of Sciences, Poland, and University of Luxembourg}
  }
\email{ wojciech.jamroga@uni.lu}

\begin{abstract}
We present \MsATL: the first tool for deciding the satisfiability 
of Alternating-time Temporal Logic (\ATL) with imperfect information.
\MsATL\ combines SAT Modulo Monotonic Theories solvers with existing \ATL\ model checkers: MCMAS and STV.
The tool can deal with various semantics of \ATL, including perfect and imperfect information,
and can handle additional practical requirements.
\MsATL\ can be applied for synthesis of games that conform to a given specification,
with the synthesised game often being minimal.
\end{abstract}

\maketitle

\section{Introduction and Motivations}
Multi-agent systems (MAS) are often treated as high-level modelling approaches for distributed and concurrent designs.
The system can be viewed as a game between the human or artificial players.
Building a formal specification of a designed system can provide various insights into the solved problem.
For instance such specifications tend to be incomplete, i.e., among its various models only selected ones are acceptable.
A minimal model conforming to the specification is most valuable:
we either obtain an implementable working example or a formally correct but non-acceptable design whose
validation may be tractable.
We are interested in synthesis of minimal game models that conform to a specification given in
\emph{Alternating-time temporal logic} (\ATL) \cite{ATL97,ATL98,ATL02,Partial-order-reductions-2018}.
Each constructive
procedure for testing satisfiability is of high practical importance, as it can be used to synthesize models from specifications.
It is utilized by various branches of computer science, including Artificial Intelligence~\cite{Koza93} and Applied Logic~\cite{Clarke81ctl,AngelisPP12}.
Model synthesis is also a field where various techniques can meet in order to attain a common goal,
such as evolutionary synthesis of correct (wrt. a formal specification) programs~\cite{KatzP17,KrawiecBSD18}.
Even if synthesis from scratch is not feasible, satisfiability-based approaches can be used in order to repair an ``almost-correct'' program~\cite{GopinathMK11,AttieCBSS15}.

\section{Theoretical Background}
{\let\thefootnote\relax\footnote{{The work of M. Kacprzak was supported by the Bialystok University of Technology, Poland, as part of the research grant WZ/WI/1/2019
of the Faculty of Computer Science and funded by the resources for research by Ministry of Science and Higher Education, Poland.
M. Knapik and W. Penczek acknowledge support from Luxembourg/Polish FNR/NCBiR project STV and CNRS/PAS project PARTIES.
}}}
\emph{Alternating-time temporal logic} (\ATL) \cite{ATL97,ATL98,ATL02} is the most important
logic for specifying strategic abilities of coalitions of agents.
\ATL\ generalizes \CTL~\cite{Clarke81ctl} by replacing the path quantifiers
$\Epath,\Apath$ with \emph{strategic modalities} $\coop{\GA}$.
Formally, the language of \ATL\ is defined by the following grammar:
$
\varphi ::= 
\prop{p}  \mid  \lnot \varphi  \mid \varphi\land\varphi  \mid
  \coop{\GA} \Next\varphi  \mid  \coop{\GA} \varphi\Until\varphi \mid \coop{\GA} \Always \varphi,
$
for $\prop{p} \in PV$ (a set of proposition variables).
Intuitively, $\coop{\GA}\gamma$ expresses that the group of agents $\GA$ has a collective strategy to enforce $\gamma$ along each path.
``$\Next$'' stands for ``next,'' ``$\Always$'' for ``always from now on,'' and $\Until$ for ``strong until.''
$\F$ (``sometimes in the future'') is defined as $\Sometm \varphi \equiv (true) \Until\varphi$.

We interpret \ATL\ over formal models of MAS.
We assume that MAS consists of $n$ agents, each assigned a set of \emph{local states},
an \emph{initial local state}, 
a set of \emph{local actions}, 
a \emph{local protocol} 
that assigns a non-empty set of available actions to each state, and a \emph{local transition function} defining possible changes of local states.
The global transition function is the composition of partial transition functions of all the agents.
To describe the interaction between agents, we have chosen Moore synchronous models \cite{ATL02}.
Moreover, for each global state, a set of propositions true in this state is defined.

A \emph{strategy} of agent $i$ is a conditional plan that specifies what $i$ is going to do in any situation.
To ensure decidability of \ATL\ model checking~\cite{Schobbens04ATL,Jamroga06atlir-eumas,Dima11undecidable},
the main technique employed by \MsATL, we focus on memoryless perfect and imperfect information strategies.
Intuitively, a memoryless imperfect information strategy for $i$ assigns a local action to each of its local states
while a perfect information strategy for $i$ assigns a local action to each global state.
Thus, perfect information strategies give agent $i$ full insight into other players' local states.
For more details see \cite{ATL02}.

The problem we are solving is to decide (in a possibly most efficient way) whether an \ATL\ formula is satisfiable.
This means, given an \ATL\ formula $\phi$, we check if there exists a model $M$
with an initial state $\iota$ in which the formula holds, i.e., $M,\iota\models\phi$.
In what follows we call this decision problem \ATLSATi\ (resp. \ATLSATI)
for imperfect (resp. perfect) information semantics of \ATL.
For more details about the theory behind \MsATL see \cite{KacprzakNiewiadomskiPenczek2020-ATLSAT}.

\section{Challenges}
\label{section:application-domain}
The main problem we are facing is a very high complexity of \ATLSATI\
and unknown complexity of \ATLSATi,
which makes non-symbolic approaches, in principle, inefficient.
The complexity of \ATLSATI\ was first proved to be EXPTIME-complete~\cite{van2003satisfiability,goranko2006complete} for a fixed number of agents and later extended to the general case in \cite{walther2006atl}.
The satisfiability of perfect information \ATL$^*$, a generalisation of perfect information \ATL, is 2EXPTIME-complete~\cite{schewe2008atl}.
The results employ techniques based on alternating tree automata.
A practically implementable tableau-based constructive decision method for \ATLSATI\
was described in \cite{GorankoShkatov2009}.
Subsequently, the tableau-based method was extended for checking ATL$^*$ \cite{david2015deciding} and ATEL
\cite{Belardinelli2014Reasoning}, an epistemic extension of \ATL\ \cite{Synt-KP04}.

Thus, there are two known methods for deciding \ATLSATI, either
by using alternating tree automata or tableau.
The first one is of rather theoretical importance. 
The tableau-based procedure has been implemented~\cite{david2015deciding}, but
it runs in 2EXPTIME and does not guarantee finding minimal models.
For \ATLSATi, it is not even known whether the problem is decidable.
A hint of its difficulty is given in~\cite{Schobbens04ATL,Jamroga06atlir-eumas} where model checking of \ATL\ with imperfect information is shown
to be $\Delta_2^P$-complete. 
No less importantly, the logic has no standard fixed-point characterisation~\cite{Bulling11mu-ijcai,Dima15fallmu}.
Note that the previous solutions are applicable only to perfect information \ATL, while the research of imperfect information \ATL\ is growing rapidly.
Our tool can deal with both variants of \ATL\ with perfect- and imperfect information.

\section{Architecture and Technology}
\label{section:usage}
\begin{figure}
  \centering
\footnotesize
\adjincludegraphics[valign=c,width=.51\columnwidth,clip=true,trim=2.3cm 12cm 5cm 8cm]{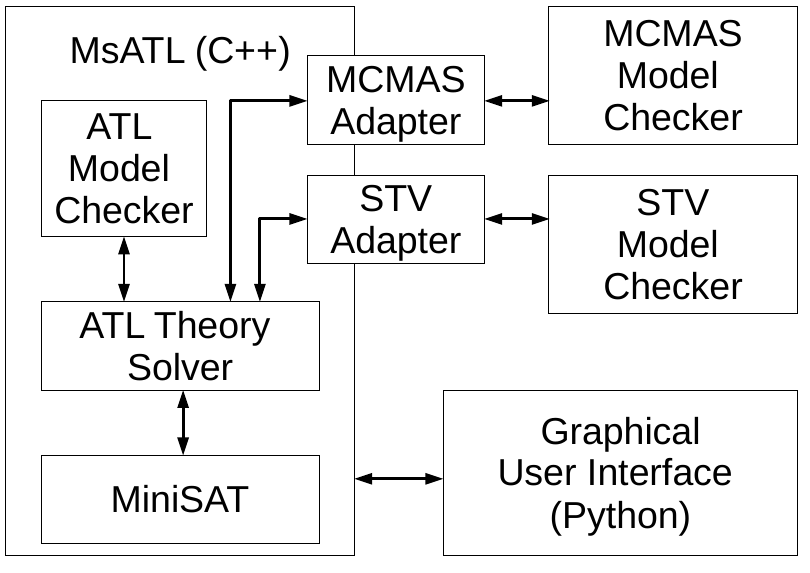}%
\hspace{0.3cm}%
\begin{tabular}{|r|r|r|r|r|}
\hline
\multicolumn{1}{|l|}{Id} & \multicolumn{1}{l|}{K} & \multicolumn{1}{l|}{C} & \multicolumn{1}{l|}{MsATL} & \multicolumn{1}{l|}{TATL} \\ \multicolumn{1}{|l|}{} & \multicolumn{1}{l|}{} & \multicolumn{1}{l|}{} & \multicolumn{1}{c|}{[sec.]} & \multicolumn{1}{c|}{[sec.]} \\
\hline
1 & 9 & 13 & 0.22 & 0.58 \\ \hline
2 & 13 & 19 & 0.23 & 6.2 \\ \hline
3 & 17 & 25 & 0.24 & 29.7 \\ \hline
4 & 20 & 31 & 0.31 & 74.6 \\ \hline
5 & 23 & 35 & 0.32 & 229.1 \\ \hline
6 & 26 & 41 & 0.34 & 551.9 \\ \hline
7 & 30 & 49 & 0.38 & 1381.5 \\ \hline
8 & 33 & 55 & 0.43 & 3947.8 \\ \hline
\end{tabular}%
\hspace{0.3cm}%
  \caption{The toolset architecture and experimental results}
  \label{fig:architecture}
\end{figure}

\MsATL\ employs a SAT solver and \ATL\ model checkers to check \ATLSAT{i/I},
and follows the concept of SAT Modulo Monotonic Theories 
solvers~\cite{Bayless-sat-modulo-2015}.
While some parts of \MsATL's architecture
(Fig.~\ref{fig:architecture}) are inspired by an earlier design
for CTL~\cite{Fast-flexible-CTL-2016},
the tackled problem is more complex,
as outlined in Sec.~\ref{section:application-domain}.
The core components of our system are:
the SAT-solver - modified MiniSAT~\cite{een2003extensible},
the \ATL\ theory solver - a module interacting with the SAT-solver,
and an \ATL\ model checker (embedded or external).
The SAT-solver is liable for manipulating 
variables
representing agents' local transitions and the valuations of propositions over global states.
The main task of the \ATL\ theory solver is to check if the current
partial valuation maintained by the SAT-solver represents a class of models that
possibly contains a model satisfying the formula.
We use external model checkers depending on the used semantics.
For memoryless perfect information strategies we use our verifier and MCMAS model-checker 
\cite{MCMAC-LomuscioRaimondi2006,MCMAS-Lomuscio2017}.
For memoryless imperfect information strategies, we use STV - the most recent tool for verification
of strategic abilities under imperfect information
\cite{STVtool-KurpiewskiJamrogaKnapik2019,STV-KurpiewskiKnapikJamroga2019,Aproximate-JamrogaKKM2019}.

\MsATL\ is modular:
we can freely attach other model checkers to expand its capabilities.
It also easily outperforms
the only other tool~\cite{david2015deciding}
over \ATLSATI\ (see Fig.~\ref{fig:architecture} (right) and Sec.~\ref{section:experimental-results}).
\MsATL\ can be used standalone
or via GUI. 
The \MsATL\ input 
requires at least: 
the number of
(1) agents,
(2) local states for each agent,
(3) proposition variables,
(4) an \ATL\ formula to be checked for satisfiability,
(5) the model checking engine.
In the case of imperfect information, a list of observable propositions for each agent is also needed.
For more details please refer to \url{http://monosatatl.epizy.com/} and video demonstration of \MsATL\ at \url{https://youtu.be/HSW-i80VEHs}.

\section{Experimental Evaluation}
\label{section:experimental-results}
Fig.~\ref{fig:architecture} (right) presents an evaluation of \MsATL performance on randomly generated\footnote{
Due to the lack of standard benchmarks for testing the satisfiability of \ATL, we have implemented
an \ATL\ formula generator.}
\ATLSATI\ instances.
\MsATL's performance is compared to the only other available
tool TATL~\cite{david2015deciding}.
The meaning of the table columns, from left to right, is as follows.
The first three contain formulas' ids;
the number of nested strategy operators;
and the total number of Boolean connectives.
Next, we present computation times of both tools, in seconds.

Table~\ref{tab:imperfect} presents experimental results for randomly generated formulae of \ATL$_{i}$ 
with \MsATL\ calling STV for the model checking subtask.
The column 'G' 
is for the number of distinct groups of agents, 
and the columns marked 'L' contain computation times (sec.) for different numbers of local states per agent.
While not comprehensive, the results show the potential of our method,
especially for some classes of \ATL\ formulae.
The experiments have been performed on Intel i5-7200U CPU/16GB Linux machine.

Satisfiability in perfect information models implies satisfiability for imperfect information, but not vice versa~\cite{Bulling14comparing-jaamas}.
To test \MsATL\ on a 
(non-randomly generated)
case that requires imperfect information, we used formula $\neg\phi$, where
$\phi \equiv \big(\neg\prop{next} \land \coop{a}\Sometm\,\prop{next} \land \coop{\emptyset}\Always(\prop{next}\rightarrow\coop{1}\Sometm\,\prop{win})\big)\ \rightarrow\ \coop{1}\Sometm\,\prop{win}$.
Intuitively, $\phi$ expresses that, if agent $a$ can get to a ``next'' state, and whenever in ``next''  
it has a follow-up strategy to win, then $a$ must also have a single strategy to win.\footnote{We could not use a more straightforward formalization, since \MsATL\ calls STV for model checking, and STV does not admit the ``nexttime'' operator $\Next$.}
Formulae like $\phi$ are known to be valid for \ATL$_{I}$ but not for \ATL$_{i}$~\cite{Bulling14comparing-jaamas}.
\MsATL\
 determined 
$\neg\phi$ to be satisfiable for \ATL$_{i}$ 
(in about 80 sec.)
and unsatisfiable for \ATL$_{I}$ (in about 11 sec.), 
which demonstrates that both functionalities of \MsATL\ are important.

\begin{table}
\caption{Satisfiability for imperfect information - the results}
\footnotesize
\begin{tabular}{|r|r|r|r|r|r|r|r|}
\hline
\multicolumn{1}{|l|}{\textbf{Id}} & \multicolumn{1}{c|}{\textbf{G}} & \multicolumn{1}{c|}{\textbf{K}} & \multicolumn{1}{c|}{\textbf{C}} & \multicolumn{1}{c|}{\textbf{L=2}} & \multicolumn{1}{c|}{\textbf{L=3}} & \multicolumn{1}{c|}{\textbf{L=4}} & \multicolumn{1}{c|}{\textbf{L=5}} \\ \hline
1 & 1 & 2 & 4 & 12.1 & 37.2 & 88.8 & 226 \\ \hline
2 & 2 & 3 & 9 & 16.4 & 52.7 & 167 & 542 \\ \hline
3 & 3 & 3 & 6 & 15.8 & 56.6 & 163 & 559 \\ \hline
4 & 3 & 4 & 6 & 22.9 & 68.1 & 194 & 746 \\ \hline
5 & 4 & 7 & 6 & 35.8 & 124 & 285 & 795 \\ \hline
6 & 5 & 13 & 13 & 70.9 & 265 & 647 & 2480 \\ \hline
7 & 5 & 17 & 15 & 88.2 & 314 & 744 & 2365 \\ \hline
8 & 5 & 21 & 18 & 106 & 383 & 1110 & 3470 \\ \hline
\end{tabular}
\label{tab:imperfect}
\end{table}

\section{Conclusions}
\label{section:conclusions}

The problem of deciding the \ATL\ satisfiability is computationally hard and the existing techniques are still not satisfactory for practical solutions.
\MsATL\ implements a novel technique, applying symbolic methods and SAT Modulo Monotonic Theories solvers for checking the \ATL\ satisfiability.
The method is universal as it can be applied to different classes of multi-agent systems \cite{LPQ10}, also with additional restrictions, and \ATL\ under various semantics. 
This is the first tool to synthesise systems under imperfect information of \ATL.
The experiments show a high potential for this approach.

\bibliographystyle{ACM-Reference-Format}

\begin{thebibliography}{00}

  
  \ifx \showCODEN    \undefined \def \showCODEN     #1{\unskip}     \fi
  \ifx \showDOI      \undefined \def \showDOI       #1{#1}\fi
  \ifx \showISBNx    \undefined \def \showISBNx     #1{\unskip}     \fi
  \ifx \showISBNxiii \undefined \def \showISBNxiii  #1{\unskip}     \fi
  \ifx \showISSN     \undefined \def \showISSN      #1{\unskip}     \fi
  \ifx \showLCCN     \undefined \def \showLCCN      #1{\unskip}     \fi
  \ifx \shownote     \undefined \def \shownote      #1{#1}          \fi
  \ifx \showarticletitle \undefined \def \showarticletitle #1{#1}   \fi
  \ifx \showURL      \undefined \def \showURL       {\relax}        \fi
  \providecommand\bibfield[2]{#2}
  \providecommand\bibinfo[2]{#2}
  \providecommand\natexlab[1]{#1}
  \providecommand\showeprint[2][]{arXiv:#2}
  
  \bibitem[\protect\citeauthoryear{Alur, Henzinger, and Kupferman}{Alur
    et~al\mbox{.}}{1997}]%
          {ATL97}
  \bibfield{author}{\bibinfo{person}{R. Alur}, \bibinfo{person}{T.~A. Henzinger},
    {and} \bibinfo{person}{O. Kupferman}.} \bibinfo{year}{1997}\natexlab{}.
  \newblock \showarticletitle{Alternating-Time Temporal Logic}. In
    \bibinfo{booktitle}{{\em Proc. of the 38th IEEE Symp. on Foundations of
    Computer Science (FOCS'97)}}. \bibinfo{publisher}{IEEE Computer Society},
    \bibinfo{pages}{100--109}.
  \newblock
  
  
  \bibitem[\protect\citeauthoryear{Alur, Henzinger, and Kupferman}{Alur
    et~al\mbox{.}}{1998}]%
          {ATL98}
  \bibfield{author}{\bibinfo{person}{R. Alur}, \bibinfo{person}{T.~A. Henzinger},
    {and} \bibinfo{person}{O. Kupferman}.} \bibinfo{year}{1998}\natexlab{}.
  \newblock \showarticletitle{Alternating-Time Temporal Logic}.
  \newblock \bibinfo{journal}{{\em LNCS\/}}  \bibinfo{volume}{1536}
    (\bibinfo{year}{1998}), \bibinfo{pages}{23--60}.
  \newblock
  
  
  \bibitem[\protect\citeauthoryear{Alur, Henzinger, and Kupferman}{Alur
    et~al\mbox{.}}{2002}]%
          {ATL02}
  \bibfield{author}{\bibinfo{person}{R. Alur}, \bibinfo{person}{T.~A. Henzinger},
    {and} \bibinfo{person}{O. Kupferman}.} \bibinfo{year}{2002}\natexlab{}.
  \newblock \showarticletitle{Alternating-Time Temporal Logic}.
  \newblock \bibinfo{journal}{{\it J. ACM}}  \bibinfo{volume}{49(5)}
    (\bibinfo{year}{2002}), \bibinfo{pages}{672--713}.
  \newblock
  
  
  \bibitem[\protect\citeauthoryear{Attie, Cherri, Dak{-}Al{-}Bab, Sakr, and
    Saklawi}{Attie et~al\mbox{.}}{2015}]%
          {AttieCBSS15}
  \bibfield{author}{\bibinfo{person}{P.~C. Attie}, \bibinfo{person}{A. Cherri},
    \bibinfo{person}{K. Dak{-}Al{-}Bab}, \bibinfo{person}{M. Sakr}, {and}
    \bibinfo{person}{J. Saklawi}.} \bibinfo{year}{2015}\natexlab{}.
  \newblock \showarticletitle{Model and program repair via {SAT} solving}. In
    \bibinfo{booktitle}{{\em 13. {ACM/IEEE} International Conference on Formal
    Methods and Models for Codesign, {MEMOCODE} 2015, Austin, TX, USA, September
    21-23, 2015}}. \bibinfo{publisher}{{IEEE}}, \bibinfo{pages}{148--157}.
  \newblock
  
  
  \bibitem[\protect\citeauthoryear{Bayless, Bayless, Hoos, and Hu}{Bayless
    et~al\mbox{.}}{2015}]%
          {Bayless-sat-modulo-2015}
  \bibfield{author}{\bibinfo{person}{S. Bayless}, \bibinfo{person}{N. Bayless},
    \bibinfo{person}{H.H. Hoos}, {and} \bibinfo{person}{A.J. Hu}.}
    \bibinfo{year}{2015}\natexlab{}.
  \newblock \showarticletitle{{SAT} Modulo Monotonic Theories}. In
    \bibinfo{booktitle}{{\em Proceedings of the Twenty-Ninth AAAI Conference on
    Artificial Intelligence}} {\em (\bibinfo{series}{AAAI'15})}.
    \bibinfo{publisher}{AAAI Press}, \bibinfo{pages}{3702--3709}.
  \newblock
  
  
  \bibitem[\protect\citeauthoryear{Belardinelli}{Belardinelli}{2014}]%
          {Belardinelli2014Reasoning}
  \bibfield{author}{\bibinfo{person}{F. Belardinelli}.}
    \bibinfo{year}{2014}\natexlab{}.
  \newblock \showarticletitle{Reasoning about Knowledge and Strategies: Epistemic
    Strategy Logic}. In \bibinfo{booktitle}{{\em Proceedings 2nd International
    Workshop on Strategic Reasoning, {SR} 2014, Grenoble, France, April 5-6,
    2014}}. \bibinfo{pages}{27--33}.
  \newblock
  
  
  \bibitem[\protect\citeauthoryear{Bulling and Jamroga}{Bulling and
    Jamroga}{2011}]%
          {Bulling11mu-ijcai}
  \bibfield{author}{\bibinfo{person}{N. Bulling} {and} \bibinfo{person}{W.
    Jamroga}.} \bibinfo{year}{2011}\natexlab{}.
  \newblock \showarticletitle{Alternating Epistemic Mu-Calculus}. In
    \bibinfo{booktitle}{{\em Proceedings of {IJCAI-11}}}.
    \bibinfo{pages}{109--114}.
  \newblock
  
  
  \bibitem[\protect\citeauthoryear{Bulling and Jamroga}{Bulling and
    Jamroga}{2014}]%
          {Bulling14comparing-jaamas}
  \bibfield{author}{\bibinfo{person}{N. Bulling} {and} \bibinfo{person}{W.
    Jamroga}.} \bibinfo{year}{2014}\natexlab{}.
  \newblock \showarticletitle{Comparing Variants of Strategic Ability: How
    Uncertainty and Memory Influence General Properties of Games}.
  \newblock \bibinfo{journal}{{\em Journal of Autonomous Agents and Multi-Agent
    Systems\/}} \bibinfo{volume}{28}, \bibinfo{number}{3} (\bibinfo{year}{2014}),
    \bibinfo{pages}{474--518}.
  \newblock
  
  
  \bibitem[\protect\citeauthoryear{Clarke and Emerson}{Clarke and
    Emerson}{1981}]%
          {Clarke81ctl}
  \bibfield{author}{\bibinfo{person}{E.M. Clarke} {and} \bibinfo{person}{E.A.
    Emerson}.} \bibinfo{year}{1981}\natexlab{}.
  \newblock \showarticletitle{Design and Synthesis of Synchronization Skeletons
    Using Branching Time Temporal Logic}. In \bibinfo{booktitle}{{\em Proceedings
    of Logics of Programs Workshop}} {\em (\bibinfo{series}{Lecture Notes in
    Computer Science})}, Vol.~\bibinfo{volume}{131}. \bibinfo{pages}{52--71}.
  \newblock
  
  
  \bibitem[\protect\citeauthoryear{David}{David}{2015}]%
          {david2015deciding}
  \bibfield{author}{\bibinfo{person}{A. David}.} \bibinfo{year}{2015}\natexlab{}.
  \newblock \showarticletitle{Deciding {ATL*} Satisfiability by Tableaux}. In
    \bibinfo{booktitle}{{\em International Conference on Automated Deduction}}.
    Springer, \bibinfo{pages}{214--228}.
  \newblock
  
  
  \bibitem[\protect\citeauthoryear{{De Angelis}, Pettorossi, and Proietti}{{De
    Angelis} et~al\mbox{.}}{2012}]%
          {AngelisPP12}
  \bibfield{author}{\bibinfo{person}{E. {De Angelis}}, \bibinfo{person}{A.
    Pettorossi}, {and} \bibinfo{person}{M. Proietti}.}
    \bibinfo{year}{2012}\natexlab{}.
  \newblock \showarticletitle{Synthesizing Concurrent Programs Using Answer Set
    Programming}.
  \newblock \bibinfo{journal}{{\em Fundam. Inform.\/}} \bibinfo{volume}{120},
    \bibinfo{number}{3-4} (\bibinfo{year}{2012}), \bibinfo{pages}{205--229}.
  \newblock
  
  
  \bibitem[\protect\citeauthoryear{Dima, Maubert, and Pinchinat}{Dima
    et~al\mbox{.}}{2015}]%
          {Dima15fallmu}
  \bibfield{author}{\bibinfo{person}{C. Dima}, \bibinfo{person}{B. Maubert},
    {and} \bibinfo{person}{S. Pinchinat}.} \bibinfo{year}{2015}\natexlab{}.
  \newblock \showarticletitle{Relating Paths in Transition Systems: The Fall of
    the Modal Mu-Calculus}. In \bibinfo{booktitle}{{\em Proceedings of {MFCS}}}
    {\em (\bibinfo{series}{Lecture Notes in Computer Science})},
    Vol.~\bibinfo{volume}{9234}. \bibinfo{publisher}{Springer},
    \bibinfo{pages}{179--191}.
  \newblock
  \showDOI{%
  \url{https://doi.org/10.1007/978-3-662-48057-1_14}}
  
  
  \bibitem[\protect\citeauthoryear{Dima and Tiplea}{Dima and Tiplea}{2011}]%
          {Dima11undecidable}
  \bibfield{author}{\bibinfo{person}{C. Dima} {and} \bibinfo{person}{F.L.
    Tiplea}.} \bibinfo{year}{2011}\natexlab{}.
  \newblock \showarticletitle{Model-checking {ATL} under Imperfect Information
    and Perfect Recall Semantics is Undecidable}.
  \newblock \bibinfo{journal}{{\em CoRR\/}}  \bibinfo{volume}{abs/1102.4225}
    (\bibinfo{year}{2011}).
  \newblock
  
  
  \bibitem[\protect\citeauthoryear{E{\'{e}}n and S{\"{o}}rensson}{E{\'{e}}n and
    S{\"{o}}rensson}{2003}]%
          {een2003extensible}
  \bibfield{author}{\bibinfo{person}{N. E{\'{e}}n} {and} \bibinfo{person}{N.
    S{\"{o}}rensson}.} \bibinfo{year}{2003}\natexlab{}.
  \newblock \showarticletitle{An Extensible SAT-solver}. In
    \bibinfo{booktitle}{{\em Theory and Applications of Satisfiability Testing,
    6th International Conference, {SAT} 2003. Santa Margherita Ligure, Italy, May
    5-8, 2003 Selected Revised Papers}} {\em (\bibinfo{series}{Lecture Notes in
    Computer Science})}, Vol.~\bibinfo{volume}{2919}.
    \bibinfo{publisher}{Springer}, \bibinfo{pages}{502--518}.
  \newblock
  
  
  \bibitem[\protect\citeauthoryear{Gopinath, Malik, and Khurshid}{Gopinath
    et~al\mbox{.}}{2011}]%
          {GopinathMK11}
  \bibfield{author}{\bibinfo{person}{D. Gopinath}, \bibinfo{person}{M.Z. Malik},
    {and} \bibinfo{person}{S. Khurshid}.} \bibinfo{year}{2011}\natexlab{}.
  \newblock \showarticletitle{Specification-Based Program Repair Using {SAT}}. In
    \bibinfo{booktitle}{{\em Tools and Algorithms for the Construction and
    Analysis of Systems - 17th International Conference, {TACAS} 2011, Held as
    Part of the Joint European Conferences on Theory and Practice of Software,
    {ETAPS} 2011, Saarbr{\"{u}}cken, Germany, March 26-April 3, 2011.
    Proceedings}} {\em (\bibinfo{series}{Lecture Notes in Computer Science})},
    Vol.~\bibinfo{volume}{6605}. \bibinfo{publisher}{Springer},
    \bibinfo{pages}{173--188}.
  \newblock
  
  
  \bibitem[\protect\citeauthoryear{Goranko and Drimmelen}{Goranko and
    Drimmelen}{2006}]%
          {goranko2006complete}
  \bibfield{author}{\bibinfo{person}{V. Goranko} {and} \bibinfo{person}{G.~Van
    Drimmelen}.} \bibinfo{year}{2006}\natexlab{}.
  \newblock \showarticletitle{Complete axiomatization and decidability of
    alternating-time temporal logic}.
  \newblock \bibinfo{journal}{{\em Theoretical Computer Science\/}}
    \bibinfo{volume}{353}, \bibinfo{number}{1-3} (\bibinfo{year}{2006}),
    \bibinfo{pages}{93--117}.
  \newblock
  
  
  \bibitem[\protect\citeauthoryear{Goranko and Shkatov}{Goranko and
    Shkatov}{2009}]%
          {GorankoShkatov2009}
  \bibfield{author}{\bibinfo{person}{V. Goranko} {and} \bibinfo{person}{D.
    Shkatov}.} \bibinfo{year}{2009}\natexlab{}.
  \newblock \showarticletitle{Tableau-based decision procedures for logics of
    strategic ability in multiagent systems}.
  \newblock \bibinfo{journal}{{\em {ACM} Trans. Comput. Log.\/}}
    \bibinfo{volume}{11}, \bibinfo{number}{1} (\bibinfo{year}{2009}),
    \bibinfo{pages}{3:1--3:51}.
  \newblock
  
  
  \bibitem[\protect\citeauthoryear{Jamroga and Dix}{Jamroga and Dix}{2006}]%
          {Jamroga06atlir-eumas}
  \bibfield{author}{\bibinfo{person}{W. Jamroga} {and} \bibinfo{person}{J. Dix}.}
    \bibinfo{year}{2006}\natexlab{}.
  \newblock \showarticletitle{Model Checking {ATL$_{ir}$} is Indeed
    {$\Delta_2^P$}-complete}. In \bibinfo{booktitle}{{\em Proceedings of
    EUMAS'06}} {\em (\bibinfo{series}{{CEUR} Workshop Proceedings})},
    Vol.~\bibinfo{volume}{223}. \bibinfo{publisher}{CEUR-WS.org}.
  \newblock
  
  
  \bibitem[\protect\citeauthoryear{Jamroga, Knapik, Kurpiewski, and
    Mikulski}{Jamroga et~al\mbox{.}}{2019}]%
          {Aproximate-JamrogaKKM2019}
  \bibfield{author}{\bibinfo{person}{W. Jamroga}, \bibinfo{person}{M. Knapik},
    \bibinfo{person}{D. Kurpiewski}, {and} \bibinfo{person}{{\L}. Mikulski}.}
    \bibinfo{year}{2019}\natexlab{}.
  \newblock \showarticletitle{Approximate verification of strategic abilities
    under imperfect information}.
  \newblock \bibinfo{journal}{{\em Artif. Intell.\/}}  \bibinfo{volume}{277}
    (\bibinfo{year}{2019}).
  \newblock
  
  
  \bibitem[\protect\citeauthoryear{Jamroga, Penczek, Dembi\'{n}ski, and
    Mazurkiewicz}{Jamroga et~al\mbox{.}}{2018}]%
          {Partial-order-reductions-2018}
  \bibfield{author}{\bibinfo{person}{W. Jamroga}, \bibinfo{person}{W. Penczek},
    \bibinfo{person}{P. Dembi\'{n}ski}, {and} \bibinfo{person}{A. Mazurkiewicz}.}
    \bibinfo{year}{2018}\natexlab{}.
  \newblock \showarticletitle{Towards Partial Order Reductions for Strategic
    Ability}. In \bibinfo{booktitle}{{\em Proceedings of the 17th International
    Conference on Autonomous Agents and MultiAgent Systems}} {\em
    (\bibinfo{series}{AAMAS '18})}. \bibinfo{pages}{156--165}.
  \newblock
  
  
  \bibitem[\protect\citeauthoryear{Kacprzak, Niewiadomski, and Penczek}{Kacprzak
    et~al\mbox{.}}{2020}]%
          {KacprzakNiewiadomskiPenczek2020-ATLSAT}
  \bibfield{author}{\bibinfo{person}{M. Kacprzak}, \bibinfo{person}{A.
    Niewiadomski}, {and} \bibinfo{person}{W. Penczek}.}
    \bibinfo{year}{2020}\natexlab{}.
  \newblock \bibinfo{title}{{SAT}-Based {ATL} Satisfiability Checking}.
  \newblock   (\bibinfo{year}{2020}).
  \newblock
  \showeprint[arxiv]{cs.LO/2002.03117}
  
  
  \bibitem[\protect\citeauthoryear{Kacprzak and Penczek}{Kacprzak and
    Penczek}{2004}]%
          {Synt-KP04}
  \bibfield{author}{\bibinfo{person}{M. Kacprzak} {and} \bibinfo{person}{W.
    Penczek}.} \bibinfo{year}{2004}\natexlab{}.
  \newblock \showarticletitle{A Sat-Based Approach to Unbounded Model Checking
    for Alternating-Time Temporal Epistemic Logic}.
  \newblock \bibinfo{journal}{{\em Synthese\/}} \bibinfo{volume}{142},
    \bibinfo{number}{2} (\bibinfo{year}{2004}), \bibinfo{pages}{203--227}.
  \newblock
  
  
  \bibitem[\protect\citeauthoryear{Katz and Peled}{Katz and Peled}{2017}]%
          {KatzP17}
  \bibfield{author}{\bibinfo{person}{G. Katz} {and} \bibinfo{person}{D. Peled}.}
    \bibinfo{year}{2017}\natexlab{}.
  \newblock \showarticletitle{Synthesizing, correcting and improving code, using
    model checking-based genetic programming}.
  \newblock \bibinfo{journal}{{\em {STTT}\/}} \bibinfo{volume}{19},
    \bibinfo{number}{4} (\bibinfo{year}{2017}), \bibinfo{pages}{449--464}.
  \newblock
  
  
  \bibitem[\protect\citeauthoryear{Klenze, Bayless, and Hu}{Klenze
    et~al\mbox{.}}{2016}]%
          {Fast-flexible-CTL-2016}
  \bibfield{author}{\bibinfo{person}{T. Klenze}, \bibinfo{person}{S. Bayless},
    {and} \bibinfo{person}{A.J. Hu}.} \bibinfo{year}{2016}\natexlab{}.
  \newblock \showarticletitle{Fast, Flexible, and Minimal {CTL} Synthesis via
    {SMT}}. In \bibinfo{booktitle}{{\em Computer Aided Verification}},
    \bibfield{editor}{\bibinfo{person}{S.~Chaudhuri} {and}
    \bibinfo{person}{A.~Farzan}} (Eds.). \bibinfo{publisher}{Springer
    International Publishing}, \bibinfo{pages}{136--156}.
  \newblock
  
  
  \bibitem[\protect\citeauthoryear{Koza}{Koza}{1993}]%
          {Koza93}
  \bibfield{author}{\bibinfo{person}{J.~R. Koza}.}
    \bibinfo{year}{1993}\natexlab{}.
  \newblock \bibinfo{booktitle}{{\em Genetic programming - on the programming of
    computers by means of natural selection}}.
  \newblock \bibinfo{publisher}{{MIT} Press}.
  \newblock
  \showISBNx{978-0-262-11170-6}
  
  
  \bibitem[\protect\citeauthoryear{Krawiec, Bladek, Swan, and Drake}{Krawiec
    et~al\mbox{.}}{2018}]%
          {KrawiecBSD18}
  \bibfield{author}{\bibinfo{person}{K. Krawiec}, \bibinfo{person}{I. Bladek},
    \bibinfo{person}{J. Swan}, {and} \bibinfo{person}{J.~H. Drake}.}
    \bibinfo{year}{2018}\natexlab{}.
  \newblock \showarticletitle{Counterexample-Driven Genetic Programming:
    Stochastic Synthesis of Provably Correct Programs}. In
    \bibinfo{booktitle}{{\em Proceedings of the Twenty-Seventh International
    Joint Conference on Artificial Intelligence, {IJCAI} 2018, July 13-19, 2018,
    Stockholm, Sweden}}. \bibinfo{publisher}{ijcai.org},
    \bibinfo{pages}{5304--5308}.
  \newblock
  
  
  \bibitem[\protect\citeauthoryear{Kurpiewski, Jamroga, and Knapik}{Kurpiewski
    et~al\mbox{.}}{2019a}]%
          {STVtool-KurpiewskiJamrogaKnapik2019}
  \bibfield{author}{\bibinfo{person}{D. Kurpiewski}, \bibinfo{person}{W.
    Jamroga}, {and} \bibinfo{person}{M. Knapik}.}
    \bibinfo{year}{2019}\natexlab{a}.
  \newblock \showarticletitle{{STV:} Model Checking for Strategies under
    Imperfect Information}. In \bibinfo{booktitle}{{\em Proceedings of the 18th
    International Conference on Autonomous Agents and MultiAgent Systems, {AAMAS}
    '19, Montreal, QC, Canada, May 13-17, 2019}}. \bibinfo{pages}{2372--2374}.
  \newblock
  
  
  \bibitem[\protect\citeauthoryear{Kurpiewski, Knapik, and Jamroga}{Kurpiewski
    et~al\mbox{.}}{2019b}]%
          {STV-KurpiewskiKnapikJamroga2019}
  \bibfield{author}{\bibinfo{person}{D. Kurpiewski}, \bibinfo{person}{M. Knapik},
    {and} \bibinfo{person}{W. Jamroga}.} \bibinfo{year}{2019}\natexlab{b}.
  \newblock \showarticletitle{On Domination and Control in Strategic Ability}. In
    \bibinfo{booktitle}{{\em Proceedings of the 18th International Conference on
    Autonomous Agents and MultiAgent Systems, {AAMAS} '19, Montreal, QC, Canada,
    May 13-17, 2019}}. \bibinfo{pages}{197--205}.
  \newblock
  
  
  \bibitem[\protect\citeauthoryear{Lomuscio, Penczek, and Qu}{Lomuscio
    et~al\mbox{.}}{2010}]%
          {LPQ10}
  \bibfield{author}{\bibinfo{person}{A. Lomuscio}, \bibinfo{person}{W. Penczek},
    {and} \bibinfo{person}{H. Qu}.} \bibinfo{year}{2010}\natexlab{}.
  \newblock \showarticletitle{Partial Order Reductions for Model Checking
    Temporal-epistemic Logics over Interleaved Multi-agent Systems}.
  \newblock \bibinfo{journal}{{\em Fundam. Inform.\/}} \bibinfo{volume}{101},
    \bibinfo{number}{1-2} (\bibinfo{year}{2010}), \bibinfo{pages}{71--90}.
  \newblock
  
  
  \bibitem[\protect\citeauthoryear{Lomuscio, Qu, and Raimondi}{Lomuscio
    et~al\mbox{.}}{2017}]%
          {MCMAS-Lomuscio2017}
  \bibfield{author}{\bibinfo{person}{A. Lomuscio}, \bibinfo{person}{H. Qu}, {and}
    \bibinfo{person}{F. Raimondi}.} \bibinfo{year}{2017}\natexlab{}.
  \newblock \showarticletitle{{MCMAS}: an open-source model checker for the
    verification of multi-agent systems}.
  \newblock \bibinfo{journal}{{\em International Journal on Software Tools for
    Technology Transfer\/}} \bibinfo{volume}{19}, \bibinfo{number}{1}
    (\bibinfo{year}{2017}), \bibinfo{pages}{9--30}.
  \newblock
  
  
  \bibitem[\protect\citeauthoryear{Lomuscio and Raimondi}{Lomuscio and
    Raimondi}{2006}]%
          {MCMAC-LomuscioRaimondi2006}
  \bibfield{author}{\bibinfo{person}{A. Lomuscio} {and} \bibinfo{person}{F.
    Raimondi}.} \bibinfo{year}{2006}\natexlab{}.
  \newblock \showarticletitle{Model checking knowledge, strategies, and games in
    multi-agent systems}. In \bibinfo{booktitle}{{\em 5th International Joint
    Conference on Autonomous Agents and Multiagent Systems {(AAMAS} 2006),
    Hakodate, Japan, May 8-12, 2006}}. \bibinfo{pages}{161--168}.
  \newblock
  
  
  \bibitem[\protect\citeauthoryear{Schewe}{Schewe}{2008}]%
          {schewe2008atl}
  \bibfield{author}{\bibinfo{person}{S. Schewe}.}
    \bibinfo{year}{2008}\natexlab{}.
  \newblock \showarticletitle{ATL* Satisfiability Is 2EXPTIME-Complete}. In
    \bibinfo{booktitle}{{\em Automata, Languages and Programming, 35th
    International Colloquium, {ICALP} 2008, Reykjavik, Iceland, July 7-11, 2008,
    Proceedings, Part {II} - Track {B:} Logic, Semantics, and Theory of
    Programming {\&} Track {C:} Security and Cryptography Foundations}}.
    \bibinfo{pages}{373--385}.
  \newblock
  
  
  \bibitem[\protect\citeauthoryear{Schobbens}{Schobbens}{2004}]%
          {Schobbens04ATL}
  \bibfield{author}{\bibinfo{person}{P.~Y. Schobbens}.}
    \bibinfo{year}{2004}\natexlab{}.
  \newblock \showarticletitle{Alternating-Time Logic with Imperfect Recall}.
  \newblock \bibinfo{journal}{{\em Electronic Notes in Theoretical Computer
    Science\/}} \bibinfo{volume}{85}, \bibinfo{number}{2} (\bibinfo{year}{2004}),
    \bibinfo{pages}{82--93}.
  \newblock
  
  
  \bibitem[\protect\citeauthoryear{van Drimmelen}{van Drimmelen}{2003}]%
          {van2003satisfiability}
  \bibfield{author}{\bibinfo{person}{G. van Drimmelen}.}
    \bibinfo{year}{2003}\natexlab{}.
  \newblock \showarticletitle{Satisfiability in alternating-time temporal logic}.
    In \bibinfo{booktitle}{{\em 18th Annual IEEE Symposium of Logic in Computer
    Science, 2003. Proceedings.}} IEEE, \bibinfo{pages}{208--217}.
  \newblock
  
  
  \bibitem[\protect\citeauthoryear{Walther, Lutz, Wolter, and Wooldridge}{Walther
    et~al\mbox{.}}{2006}]%
          {walther2006atl}
  \bibfield{author}{\bibinfo{person}{D. Walther}, \bibinfo{person}{C. Lutz},
    \bibinfo{person}{F. Wolter}, {and} \bibinfo{person}{M. Wooldridge}.}
    \bibinfo{year}{2006}\natexlab{}.
  \newblock \showarticletitle{ATL satisfiability is indeed EXPTIME-complete}.
  \newblock \bibinfo{journal}{{\em Journal of Logic and Computation\/}}
    \bibinfo{volume}{16}, \bibinfo{number}{6} (\bibinfo{year}{2006}),
    \bibinfo{pages}{765--787}.
  \newblock
  
  
  \end{thebibliography}

\end{document}